\documentclass[runningheads]{llncs}
\usepackage[T1]{fontenc}
\usepackage{graphicx}
\usepackage{soul}
\usepackage{color}
\usepackage{subcaption}
\usepackage{bbm}
\usepackage{amsmath}
\usepackage{amssymb}
\usepackage{pifont}%
\usepackage{booktabs}
\newcommand{\cmark}{\ding{51}}%
\newcommand{\xmark}{\ding{55}}%
\usepackage{tcolorbox}

\begin{document}

\title{LIBR+: Improving Intraoperative Liver Registration by Learning the Residual of Biomechanics-Based Deformable Registration}
%
%
\author{Dingrong Wang\inst{1} \and
Soheil Azadvar\inst{1} \and
Jon Heiselman\inst{2} \and 
Xiajun Jiang\inst{1} \and
Michael Miga\inst{2} \and
Linwei Wang\inst{1} }
%
\authorrunning{Wang et al.}

\institute{Rochester Institute of Technology, Rochester NY 14623, USA \and
Vanderbilt University, Nashville TN, USA}

\maketitle              
\begin{abstract}
The surgical environment imposes unique challenges to 
the intraoperative registration of organ shapes to their preoperatively-imaged geometry.
Biomechanical model-based registration remains popular, while deep learning solutions remain limited due to the sparsity and variability of intraoperative measurements and the limited 
ground-truth deformation of an organ that can be obtained during the surgery. 
In this paper, 
we propose a novel 
\textit{hybrid} registration approach that
leverage a linearized iterative boundary reconstruction (LIBR) method based on linear elastic biomechanics, 
and use deep neural networks to learn its residual to the ground-truth deformation (LIBR+). 
We further formulate a dual-branch 
spline-residual graph convolutional neural network (SR-GCN) 
to assimilate information from sparse and variable intraoperative measurements and effectively propagate it through the geometry of the 3D organ. 
Experiments on a large intraoperative liver registration dataset demonstrated the consistent improvements achieved by LIBR+ in comparison to existing rigid, 
biomechnical model-based non-rigid, 
and deep-learning based non-rigid 
approaches to intraoperative liver registration.

\end{abstract}
\section{Introduction}
Surgical resection of liver tumor 
relies on an accurate intraoperative localization of 
interventional targets and critical anatomy 
residing 
beneath the liver surface. 
However, 
the liver experiences significant deformations during surgery,  
compromising
the integrity of preoperative surgical plans established from preoperative images \cite{Heiselman_JMI18}. Surgical navigation systems, which depend on image-to-physical registration to align anatomical models and surgical plans into the intraprocedural coordinate frame, requires an ability to infer soft-tissue deformations of the  
preoperatively-imaged liver anatomy using 
sparse intraoperative measurements. 

Various image-to-physical registration 
methods have been proposed to enable image-guided liver surgery.
Rigid registration is a routine technique that is fast to compute, where several advances have been made to improve robustness to data quality \cite{Clements_MP08,Yang_IJCARS23}. However, the presence of intraoperative soft tissue deformations limits its applicability to image-guided liver surgery. For deformable registration, biomechanical model-based approaches 
have been a popular choice \cite{Brock_MP05,Suwelack_MP11,Plantefeve_ABME16,Mestdagh_MICCAI22,Heiselman_JMI24}. 
These approaches 
use governing biomechanical  equations to constrain the deformation of an organ that is optimized iteratively to match intraoperative sparse measurements. 
Despite substantial success, however, 
tradeoffs often have to be made to balance computational efficiency and model complexity to meet intraoperative time constraints and resources. Simplifying assumptions for soft tissue deformation, such as linear elastic models of underlying biomechanics, thus remain prevalent in state-of-the-art algorithms 
\cite{Heiselman_JMI24}.
These simplifying assumptions 
lead to a gap in the ability for these models to match real-world deformations 
and thus limits the registration accuracy that can be achieved \cite{Marchesseau17}.

In the meantime, 
while deep-learning registration methods have flourished 
\cite{xiao2023deep}, 
their penetration to image-guided liver surgery has been limited. 
A fundamental challenge arises from the difficulty to obtain a diverse and large quantity of ground-truth (GT) deformations, \textit{i.e.}, the labels, in the surgical environment. The sparsity and variety of 
measurement data from the intraoperative phase and 
the inter-subject variations in liver geometry 
further adds significant challenges. 
In \cite{jia2021improving}, point cloud reconstruction models are used to improve inference of the intraoperative deformation of the liver.
In \cite{pfeiffer2020non} (V2S-Net), 
a encoder-decoder architecture with 3D CNN is used to learn a deformation model using the preoperative liver surface mesh and intraoperative liver surface sparse points. 
These methods rely heavily on the availability of GT deformations as supervision, which is limited in practice. 
As a result, their performance in a recent \textit{sparse data challenge} \cite{brewer2019image,Heiselman_JMI24} on intraoperative liver registration was not competitive compared to biomechanical model-based approaches. 
How to address limited training labels, 
sparse and variable input data, 
and the topology and inter-/intra-subject variations of anatomical organs remain open challenges in deep learning solutions to image-to-physical registration \cite{zou2022review,fu2020deep,xiao2023deep}.

\begin{figure}[!t]
    \centering
    \includegraphics[width=.9\textwidth]{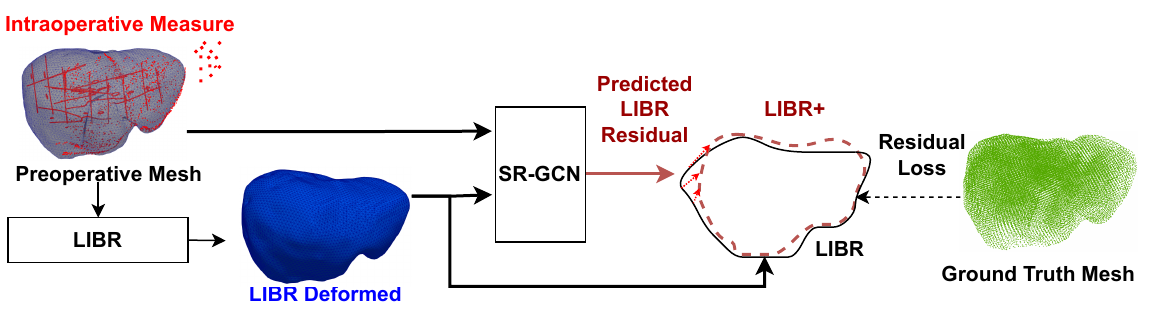}
    \caption{Overview of LIBR+}
    \label{fig: diagram}
\end{figure}

In this paper, we propose to overcome these challenges  via a novel \textit{hybrid} image-to-physical registration approach that combines the advantages of biomechanical model-based and deep-learning methods. We leverage a linearized iterative boundary reconstruction (LIBR) method based on linear elastic biomechanics \cite{Heiselman_TMI20}, and use deep neural networks to learn its residual to the GT deformation. 
We denote it as LIBR+ which, 
as outlined in Fig \ref{fig: diagram}, 
has two key innovations. 
First, compared to a purely data-driven approach to learn 
all aspects of soft-tissue biomechanics underlying GT deformations, 
the proposed 
residual-learning formulation  
removes the need for the network to recapitulate the known linear elastic components of
soft-tissue biomechanics. Additionally, 
by learning LIBR residuals corresponding to different intraoperative data configurations, 
it lowers the requirement on 
the limited GT deformation that can be obtained in surgical requirements.
Second, 
we realize this residual learning by formulating 1) spline based graph convolutional neural network (spline-GCN) \cite{fey2018splinecnn} defined over the liver mesh,
and 2)
its interaction with 
intraoperative sparse measurements through convolutions over bipartite graphs.
This spline-residual GCN (SR-GCN) architectural design 
is able to accommodate different liver meshes, 
as well as different configurations of sparse intraoperative measurements. 

We evaluated LIBR+ on a large liver registration dataset used in \cite{Heiselman_TMI20} and \cite{Heiselman_TMI21}, which contains 10 liver models of preoperative and GT tetrahedron meshes, as well as over 7,000 configurations of sparse surface and subsurface measurements of the GT liver geometries.
This dataset also includes baseline registration results of a rigid salient feature weighted iterative closest point (wICP) algorithm \cite{Clements_MP08} and 
LIBR 
\cite{Heiselman_TMI20}, 
which we used to compare with LIBR+ along with 
a deep-learning based nonlinear registration model V2S \cite{pfeiffer2020non}. 
Comparison results and ablative studies demonstrated consistent improvements of LIBR+ 
over both biomechanical model-based and data-driven baselines in intraoperative liver registration.




\section{Background}

\subsubsection{LIBR:}
Briefly, the LIBR method uses linear elastic biomechanics to establish a superposed deformation basis that can be used to rapidly reconstruct changes in the soft-tissue deformation state between the preoperative organ geometry and its intraoperative conformation. A linear elastic finite element model of the organ geometry is first constructed, and then control points are uniformly spaced over the organ surface and perturbed in each Cartesian direction to establish deformation responses that decompose organ motions into a series of spatially localized deformation functions. The algebraic span of these localized deformation response functions produces an orthogonal, linear elastic deformation basis that can be resolved through an iterative estimation process, to solve for a vector of basis weights that maximizes agreement between the total linear elastic deformation response and sparse observations of the intraoperative organ shape. The reader is directed to reference~\cite{Heiselman_TMI20} for comprehensive details. LIBR+ will use the LIBR output as a linear-physics prior for refining the resulting deformation function and recapturing the nonlinear physical effects that will more accurately model soft tissue deformations within intraoperative runtime constraints.

\subsubsection{Spline-GCNN:} 
Spline-GCNN is a variant of deep neural networks for irregular structured and geometric input, \textit{e.g.}, graphs or meshes. For each channel $l$ of a feature map, the spline convolution kernel is defined as: $ g_l(\mathbf{u}) = \sum_{\mathbf{p} \in \mathcal{P}} w_{\mathbf{p}, l} B_{\mathbf{p}}(\mathbf{u})$, 
where $1 \leq l \leq C$ and $C$ is the number of channels. 
$B_{\mathbf{p}}(\mathbf{u})$ denotes 
the 
spline basis  
and $w_{\mathbf{p}, l}$ are trainable parameters. 
Since the B-spline basis is conditioned on local geometry, the learned kernel can be applied across graphs and the convolution incorporates geometrical information within the graph. 
This allow us to train and test across different liver meshes, 
 as well as to accommodate 
sparse measurements from different modalities with
different size and structure.






\section{Methodology}

\begin{figure}[!t]
    \centering
    \includegraphics[width=\textwidth]{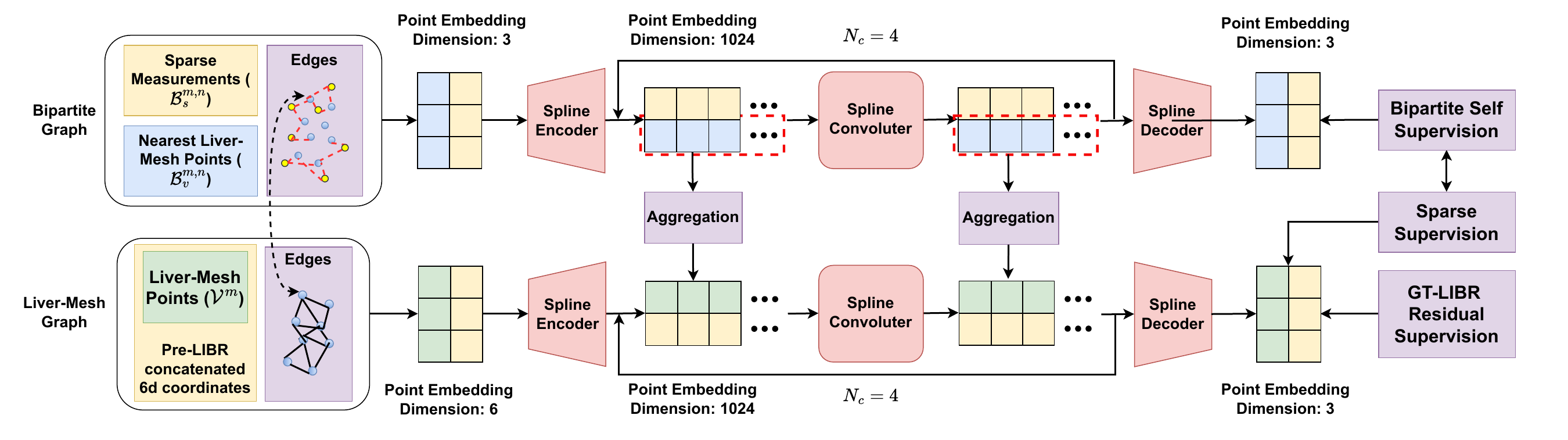}
    \caption{Outline of SR-GCN architecture.}
    \label{fig:splineED}
\end{figure}

Consider a preoperative liver anatomy $\mathcal{I}$ in the form of a tetrahedral mesh and intraoperative sparse measurement data $\mathcal{S}$, we obtain the corresponding LIBR-deformed liver mesh $\mathcal{L}$ corresponding to $\mathcal{S}$.
The goal of LIBR+ is then to take 
$\mathcal{I}, \mathcal{L}$ and $\mathcal{S}$
as inputs and predict 
the residual in $\mathcal{L}$
to the GT liver mesh $\mathcal{T}$:
\begin{equation}
    Residual(\mathcal{L}, \mathcal{T}) = \mathcal{F} (\mathcal{I}, \mathcal{L}, \mathcal{S})
\end{equation}
On each pair of preoperative and GT deformed liver mesh $\{ \mathcal{I}^m,  \mathcal{T}^m \}_{m=1}^M$, 
there can be multiple LIBR input-output pairs $ \{ \mathcal{S}^{m,n},  \mathcal{L}^{m,n} \}_{n=1}^N$ that can be used for LIBR+ to learn LIBR residuals to GT deformation, 
where $M$ is the total number of unique liver models and $N$ the total number of instances of sparse measurements available for LIBR solutions.
As illustrated in Fig.~\ref{fig:splineED}, 
function $\mathcal{F}$ in LIBR+ is realized with two major backbone branches:
graph convolution on the liver mesh (bottom branch), and 
its assimilation of intraoperative measurements $\mathcal{S}$ (top branch). 
Below we describe in detail constructions of the graphs as the domain on which $\mathcal{F}$ is operated (Section \ref{subsec:graph}), 
architectural design of $\mathcal{F}$ (Section \ref{subsec:encdec}), 
and the final objective function to optimize $\mathcal{F}$ (Section \ref{subsec:loss}).


\subsection{Construction of bipartite and liver-mesh graphs }
\label{subsec:graph}

To take into account the geometric topology of the liver and its spatial relation to intraoperative measurements, 
we construct two graphs to support LIBR+.

\subsubsection{Liver-mesh graph:} All liver meshes $\mathcal{I}^m, \mathcal{T}^m$ and $\mathcal{L}^{m,n}$ share the same tetrahedron topology  
with $N^{m}_{p}$ number of vertices and $N^{m}_{e}$ number of edges
but have different coordinate values on the vertices. 
We thus construct a graph adopting the vertices $\mathcal{V}^m$ and edges $\mathcal{E}^m$ of the tetrahedron mesh. 
For each input liver-mesh graph to $\mathcal{F}$, the vertex attributes $\mathbf{V}^{m,n} \in \mathbb{R}^{N^{m}_p \times 6}$ are defined as the concatenation of coordinates from preoperative and LIBR-deformed liver meshes, 
and the edge attributes $\mathbf{U}^{m,n} \in [0, 1]^{N^{m}_p \times N^{m}_p \times 1}$ are defined as pseudo indexes ranging from 0 to 1.
We denote this liver-mesh graph as 
$\mathcal{G}^{m,n} = (\mathcal{V}^m, \mathcal{E}^m, \mathbf{V}^{m,n}, \mathbf{U}^{m,n})$.




\subsubsection{Bipartite graph:} To allow information from intraoperative measurements to propagate to the liver-mesh graph, 
we define a bipartite graph between the sets of intraoperative measurement points $\mathcal{B}_s^{m,n}$ and their
\textit{corresponding}
points $\mathcal{B}_v^{m,n}$ on the liver mesh. 
To establish this correspondence,  
for each point in the set of $N_s$ number of points in $\mathcal{B}_s^{m,n}$, 
we identify its $N_{top}$ nearest neighbor points from the GT deformed mesh $\mathcal{T}^m$ 
to form a 
set of $N_{v}$ number of 
vertices $\mathcal{B}_{v}^{m,n}$ 
on the liver-mesh graph $\mathcal{G}^{m,n}$.
We then
construct a bipartite graph  
between these two sets of vertices as
$\mathcal{G}^{m,n}_{bipar} = \{\mathcal{B}_s^{m,n},\mathcal{B}_{v}^{m,n}, \mathcal{E}_b^{m,n}, 
\mathbf{B}_s^{m,n}, 
\mathbf{B}_{v}^{m,n},
\mathbf{U}_b^{m,n}
\}$, where 
the 
edge connection $\mathcal{E}_b^{m,n}$ 
between $\mathcal{B}_s^{m,n}$ and $\mathcal{B}_v^{m,n}$ are defined by the $N_{top}$ nearest neighbors, 
with edge attributes 
$\mathbf{U}_b^{m,n} \in [0, 1]^{N_s \times N_s \times 1}$
defined similarly to $\mathbf{U}^{m,n}$ above. 
The  
vertex attributes $\mathbf{B}_s^{m,n} \in \mathbb{R}^{N_s \times 3}$ 
are defined by the 3D coordinates of the intraoperative measurement points, 
and $\mathbf{B}_{v}^{m,n} \in \mathbb{R}^{N_{v} \times 3}$ the 3D coordinates of the preoperative liver-mesh. 
No edge connection exists within $\mathcal{B}_s^{m,n}$ or $\mathcal{B}_v^{m,n}$.

\subsection{Spline Residual Graph Convolutional Networks (SR-GCN)}
\label{subsec:encdec}

Given the liver-mesh graph 
$\mathcal{G}^{m,n}$ and the bipartite grahp $\mathcal{G}_{bipar}^{m,n}$, 
function $\mathcal{F}$ is realized via spline-GCNN with two primary branches as illustrated in 
Figure~\ref{fig:splineED}. 
Each branch 
consists of a spline encoder 
(
$f \in \mathbb{R}^{V, 6}\rightarrow \mathbb{R}^{V, 1024})$, 
convoluters 
($\mathit{g}\in \mathbb{R}^{V, 1024}\rightarrow \mathbb{R}^{V, 1024}$),
and decoder 
($\mathit{h} \in \mathbb{R}^{V, 1024}\rightarrow \mathbb{R}^{V, 3}$), 
with aggregation of information in between 
($V$ denotes the number of vertices in either graph).



\subsubsection{Bipartite branch}: 
On the bipartite graph, 
the spline encoder $\mathit{f}^b_{\theta_b}$ 
transforms 
vertex attributes $\{\mathbf{B}_s^{m,n}, \mathbf{B}_{v}^{m,n}\}$ into  
feature maps $\mathbf{x}^{l=0}_{b,s}, \mathbf{x}^{l=0}_{b,v}$ which, 
through $N_{c}$ times of spline convoluter $\mathit{g}^b_{\phi_b}$, 
delivers vertex feature maps $\mathbf{x}^{l=N_{c}}_{b,s}, \mathbf{x}^{l=N_{c}}_{b,v}$ 
that is 
decoded to 
$\mathbf{y}_{b,s}, \mathbf{y}_{b,v}$ through the 
spline decoder $\mathit{h}_{\psi_b}$: 
\begin{align}
\label{eq: encoder_b}
\{ \mathbf{x}^{l=0}_{b,s}, \mathbf{x}^{l=0}_{b,v} \} & = \mathit{f}^b_{\theta_b}(\{\mathbf{B}_s^{m,n}, \mathbf{B}_{v}^{m,n}\}) \\
\label{eq: convoluter_b}
\{ \mathbf{x}^{l+1}_{b,s}, \mathbf{x}^{l+1}_{b,v} \}  & = \mathit{g}^b_{\phi_b}(\{\mathbf{x}^{l}_{b,s}, \mathbf{x}^{l}_{b,v}\}), \quad l=0:N_c-1 \\
\label{eq: decoder_b}
\mathbf{y}_b  = \{ \mathbf{y}_{b,s}, \mathbf{y}_{b,v} \}  &=  \mathit{h}^b_{\psi_b}(\{\mathbf{x}^{l=N_{c}}_{b,s}, \mathbf{x}^{l=N_{c}}_{b,v}\})
\end{align}
where $\theta_b, \phi_b$ and $\psi_b$ parameterize the neural networks, and $\mathbf{y}_b$ is intended to reconstruct the 
input vertex attributes $\{\mathbf{B}_s^{m,n}, \mathbf{B}_{v}^{m,n}\}$.

\subsubsection{Liver-mesh branch:} 
Functions on the liver-mesh graph follow a similar composite of operations starting with the vertex attribute $\mathbf{V}^{m,n}$, with a key difference that 
at each convolution layer $l$, the convoluter $\mathit{g}_\phi$'s input $\mathbf{x}^{l-1}_{v}$ will be aggregated (averaged) with 
the feature map $\mathbf{x}^{l-1}_{b,v}$ obtained from the bipartite branch on the corresponding graph vertices
through an \textit{Aggregation} module:
\begin{align}
\label{eq: encoder}
\mathbf{x}^{l=0}_{v} & = \mathit{f}_\theta(\mathbf{V}^{m,n}), \quad 
\mathbf{x}^{l+1}_{v}   = \mathit{g}_\phi(\mathbf{x}^{l}_{v}), \quad 
\mathbf{x}^{l}_{v} = \frac{1}{2} (\mathbf{x}^{l}_{v} + \tilde{\mathbf{x}}^{l}_{b,v}), \quad  
l=0:N_{c}-1
\\
\label{eq: decoder}
\mathbf{y}_v & = 
\mathit{h}_\psi(
\mathbf{x}^{l=N_{c}}_{v}) 
\end{align}
where $\theta, \phi$ and $\psi$ parameterize the neural networks. 
$\tilde{\mathbf{x}}^{l}_{b,v}$ consists of  
$\mathbf{x}^{l}_{b,v}$ on 
vertices $\mathcal{B}_v^{m,n}$ corresponding to sparse measurement points, and 0's elsewhere. 

\subsection{Optimization of LIBR+}
\label{subsec:loss}

The SR-GCN is trained in two stages. 
We first optimize the 
bipartite branch using a self-supervised reconstruction loss. 
With this branch fixed, 
we optimize the liver-mesh branch to predict the residual of the LIBR solutions on the entire liver mesh, subject to the constraints on sparse measurement points. 


The bipartite branch of SR-GCN is optimized by a L2-reconstruction loss:
\begin{align}
\label{eq: bipar}
\mathcal{L}_{b}(\theta_b,\phi_b,\psi_b) = & \frac{1}{M*N}\sum_{m=1}^{M}\sum_{n=1}^{N} \left[\{\mathbf{B}_s^{m,n}, \mathbf{B}_{v}^{m,n},\}- \mathbf{y}_b \right]^2
\end{align}

The liver-mesh graph branch is then trained to predict the residual of each LIBR solution $\mathbf{L}_v^{m,n}$ to the GT deformation $\mathbf{T}_v^m$:
\begin{align}
\label{eq: residual}
\mathcal{L}_{residual}(\theta, \phi, \psi) = \frac{1}{M*N}\sum_{m=1}^{M}\sum_{n=1}^{N}\left[(\mathbf{T}^m-\mathbf{L}^{m,n})- \mathbf{y}_v\right]^2
\end{align}
with an additional supervision on vertices $\mathcal{B}_v^{m,n}$ by sparse measurement points:

\begin{align}
\label{eq: sparse}
\mathcal{L}_{sparse}(\theta, \phi, \psi) = \frac{1}{M*N}\sum_{m=1}^{M}\sum_{n=1}^{N} \left[\mathbf{B}_{s}^{m,n} - \mathbf{L}^{m,n}_{v} (\mathcal{B}_v^{m,n}) - \mathbf{y}_{v} (\mathcal{B}_v^{m,n}) \right]^2
\end{align}
with a final loss as $\mathcal{L}_{m} = \mathcal{L}_{residual} + \lambda \mathcal{L}_{sparse}$ balanced by hyperparaemter $\lambda$.


\section{Experiments and Results}

\subsubsection{Data:} 
We utilized the 
liver deformation data set 
described in \cite{Heiselman_TMI20} and \cite{Heiselman_TMI21}, which 
contain 7,000 total registration scenarios across ten unique liver deformations. Each of the 700 registration scenarios per deformation represent widely varying levels of sparsity of intraoperative data, from which we used sparse surface data only, sparse surface data plus sparse subsurface data from 1-3 iUS planes, and sparse surface data plus dense subsurface data from 16 iUS planes. 
The base dataset in \cite{Heiselman_TMI20} included nine GT deformations derived from three unique patient-specific liver geometries (1/2/3), interposed with three different deformation patterns (L/R/W) derived from physical simulations of intraoperative laparoscopic surgery. A tenth dataset (4-V) in \cite{Heiselman_TMI21}, which introduces a fourth unique liver geometry and a deformation pattern derived from physical simulation of open surgery, was also included for testing only. These datasets included 1) the preoperatively-imaged liver geometry, 2) the GT liver geometry caused by intraoperative deformations, and 3) sparse point cloud data of intraoperative measurements of the anterior surface and subsurface features 
of the liver.  Rigid wICP 
and deformable LIBR registration results
were also included. 
Because the original meshes had $\sim$30K vertices, 
we coarsened them
with  
the \textit{iso2mesh} library in Matlab.
We also coarsened the sparse measurements 
with a one-shot point-cloud down-sampling algorithm in Matlab, with carefully chosen hyper-parameters to preserve its geometry and its ratio to the liver meshes.  


\subsubsection{Baselines and metrics:} 
We considered three baselines representing three categories of image-to-physical registration approaches: 
1) rigid registration by wICP \cite{Clements_MP08}, 
2) deformable LIBR registration 
\cite{Heiselman_TMI20}, 
and 3) DNN based nonlinear registration model V2S \cite{pfeiffer2020non}. 
We considered a quantity metric of TRE
as L2 distance between LIBR+ deformed mesh and GT deformed mesh. 

\begin{table}[!t]
  \centering
  \caption{Comparison of TRE (mm) results in different training-test splits.}
    \begin{tabular}{ccccc}
    \toprule
     & \multicolumn{4}{c}{Hold-Out Test Split} \\
\cmidrule{2-5} Method & Random & Geometry-Deformation &Geometry & 4-V \\
    \midrule
    wICP \cite{Clements_MP08} & 13.005$\pm$3.118 & 13.865$\pm$2.133 & 14.254$\pm$1.626 & 15.352$\pm$0.692  \\
    LIBR \cite{Heiselman_TMI20} & 5.915$\pm$0.719& 6.314$\pm$2.048 & 6.668$\pm$2.754 & 6.706$\pm$1.804  \\
    V2S \cite{pfeiffer2020non}  & 5.037$\pm$0.169  & 6.548$\pm$1.732 & 6.980$\pm$0.433 & 8.804$\pm$0.226 \\
    \textbf{LIBR+}  & \textbf{3.234$\pm$0.760}  & \textbf{3.159$\pm$0.080} & \textbf{3.365$\pm$0.850}  & \textbf{3.702$\pm$1.170} \\
    \bottomrule
    \end{tabular}
  \label{tab: baseline}
\end{table}

\subsubsection{Comparison with Baselines:}
We evaluated LIBR+ and its baselines on multiple different training-test splits of the nine geometry-deformation pattern combinations, in increasing difficulty as:
1) a random split regardless of geometry-deformation combinations. 
2) a split leaving geometry-deformation combinations (1-L, 1-R, and 3-W) out, 
and 3) three splits that each leaves one geometry (including all its deformation patterns) out.  
We further did a test where the models are trained on all nine geometry-deformation combinations, and tested on 4-V.


Tables ~\ref{tab: baseline} summarizes TRE results on the different splits, showing significant improvements of LIBR+ in comparison to all baselines across all splits.  
Notably,  
while all model's performance in general decreased as 
the generalization difficulty increased over the splits, 
LIBR+ 
demonstrated a significantly higher robustness 
(17\% performance drop comparing worst to best performance) compared to V2S (75\% drop) -- 
this highlights the challenge of over-fitting for pure data-driven solutions in this problem,  
and the ability of LIBR+ to overcome this challenge.



\begin{figure*}[t]
     \centering
\includegraphics[width=\linewidth]{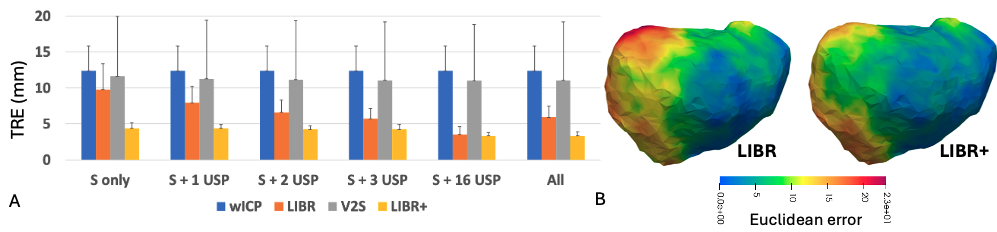}
     \caption{A: TRE by categories of sparsae measurements used. S = Surface; USP = US Plane. B: Visual examples of vertex-wise error maps from LIBR \textit{vs}.\ LIBR+.} 
      \label{fig: visual}
 \end{figure*}

Figure~\ref{fig: visual}A summarizes TRE results across all splits, 
showing the improvements of LIBR+ over baselines given different categories of intraoperative measurements.
Figure~\ref{fig: visual}B provides visual TRE examples of LIBR \textit{vs}.\ LIBR+.    

\subsubsection{Ablation Results:}
We further ablated the SR-GCN on the random data split,  
considering combinations of inputs and losses 
as detailed in  Table~\ref{tab: ablative}. 
As shown, the use of residual-loss as supervision ($\mathcal{L}_{residual}$) 
was beneficial compared to the use of GT deformation ($\mathcal{L}_{gt}$) as supervision (2 \textit{vs}.\ 1 or 6 \textit{vs}. 5). 
The addition of LIBR as input, 
the bipartite branch, 
and the constraints on sparse measurement points 
$\mathcal{L}_{sparse}$ (3, 4, and 5) further in turn improved the registration results.

\begin{table}[!t]
  \centering
  \caption{Ablation results on random data split.}
    \begin{tabular}{cccccccc}
    \toprule
 Setting  & \multicolumn{3}{c}{Input Condition} & \multicolumn{3}{c}{Loss Condition} & Metric\\
\midrule
  &  Pre &  LIBR & Bipartite & $L_{gt}$ & $L_{residual}$ & $L_{sparse}$ & TRE \\
    \midrule
 1 & \cmark & \xmark & \xmark & \cmark & \xmark & \xmark & 4.215$\pm$0.495 \\
 2 & \cmark & \xmark & \xmark & \xmark & \cmark & \xmark & 3.857$\pm$0.553 \\
 3 & \cmark  & \cmark & \xmark & \xmark & \cmark & \xmark & 3.985$\pm$0.795 \\
 4 & \cmark  & \cmark & \cmark & \xmark & \cmark & \xmark & 3.346$\pm$0.831 \\
 5 & \cmark  & \cmark & \cmark & \cmark & \xmark & \cmark & 3.325$\pm$0.784 \\
 \textbf{6 (Ours)} & \cmark  & \cmark & \cmark & \xmark & \cmark & \cmark & \textbf{3.234$\pm$0.763}\\
\bottomrule
\end{tabular}
\label{tab: ablative}
\end{table}



\section{Conclusion}
In this work, 
we propose LIBR+ as a novel framework to improve intraoperative liver registration by learning the residual of a biomechnical model-based method LIBR. 
Quantitative and qualitative experiments demonstrated the significant improvements achieved by LIBR+ compared to its biomechanical model-based and data-driven baselines, especially in its robustness to generalize to unseen geometry and deformation patterns. 
Future work will explore improving LIBR+ by adding skip connections between the LIBR-deformed mesh to the output of LIBR+, 
as well as embedding the SR-GCN in the latent space of a graph pooling-unpooling architecture to realize registrations on the fully-resolved liver meshes. 
Extension of LIBR+ to other image-guided surgery will also be investigated.


\bibliographystyle{splncs04}
\bibliography{main}

\newpage
\appendix

\begin{center}
   {\large \bf  Supplementary Material}
\end{center}

\begin{figure}[htbp]
    \centering
    \begin{subfigure}{0.32\linewidth}
        \includegraphics[width=\linewidth]{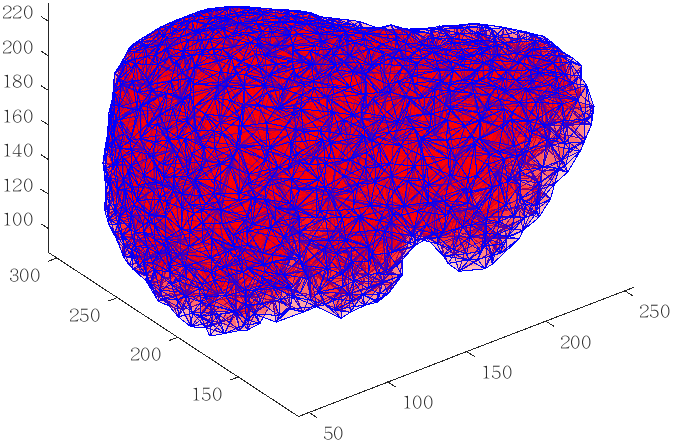}
        \caption{preoperative mesh of liver model 1 without before left forces applied.}
    \end{subfigure}
    \hfill
    \begin{subfigure}{0.32\linewidth}
        \includegraphics[width=\linewidth]{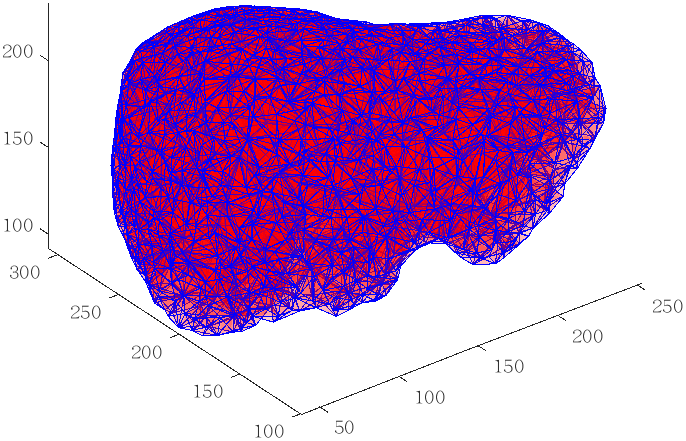}
        \caption{LIBR-deformed mesh of liver model 1-L without before left forces applied.}
    \end{subfigure}
    \hfill
    \begin{subfigure}{0.32\linewidth}
        \includegraphics[width=\linewidth]{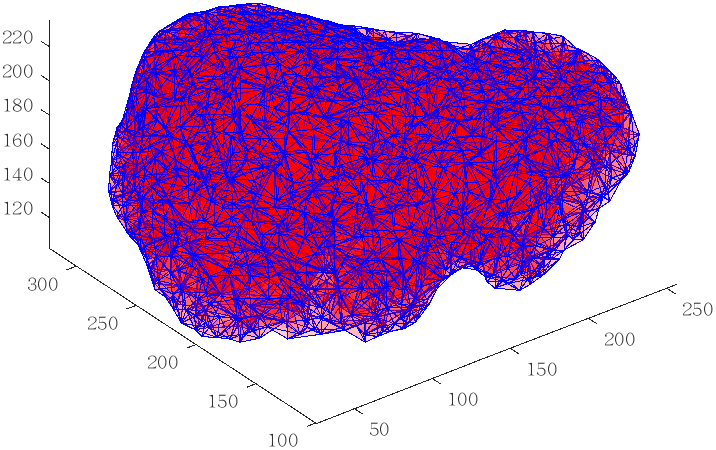}
        \caption{GT deformed mesh of liver model 1-L after left forces applied.}
    \end{subfigure}
    \caption{Similar topology of preoperative (a), LIBR (b) and GT (c) deformed mesh after data coarsening. This gives a verified support of the assumption that a shared tetrahedron topology could be extracted from these three meshes after data coarsening.}
    \label{fig: verify} 
\end{figure}

\begin{figure}[htbp]
    \centering
    \begin{subfigure}{0.32\linewidth}
        \includegraphics[width=\linewidth]{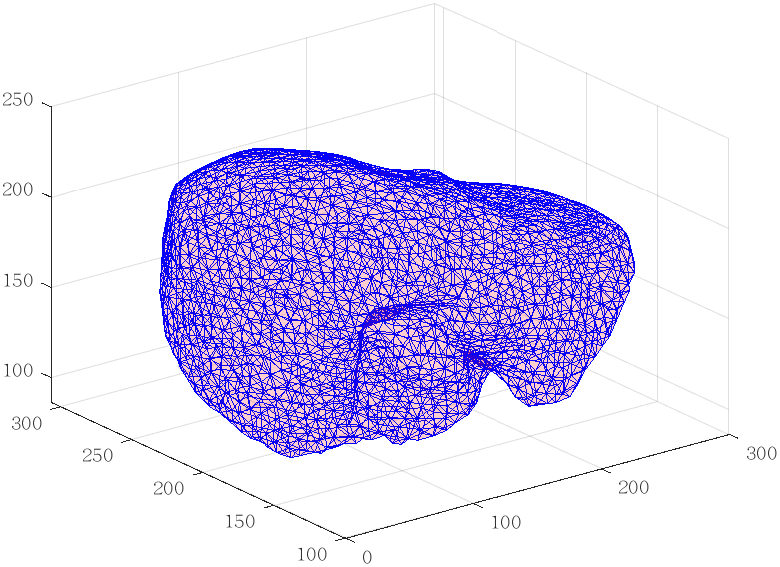}
        \caption{Surface Detection}
    \end{subfigure}
    \hfill
    \begin{subfigure}{0.32\linewidth}
        \includegraphics[width=\linewidth]{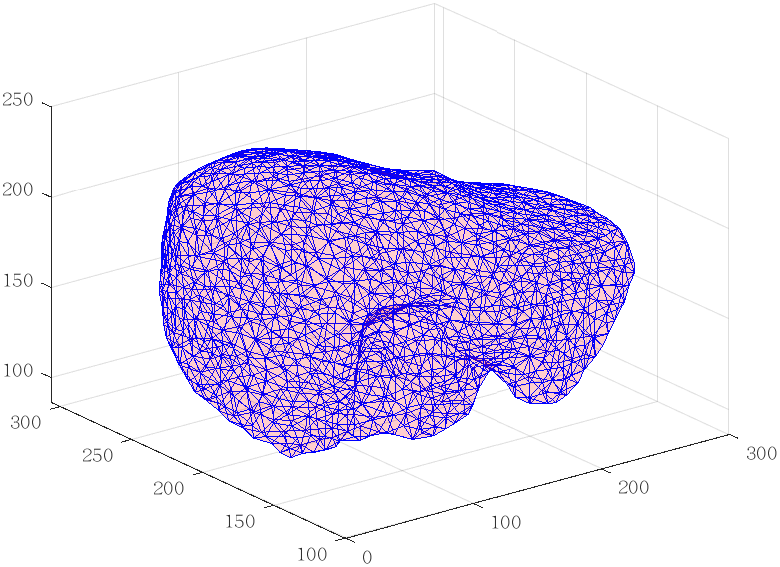}
        \caption{Surface Down-Sampling}
    \end{subfigure}
    \hfill
    \begin{subfigure}{0.32\linewidth}
        \includegraphics[width=\linewidth]{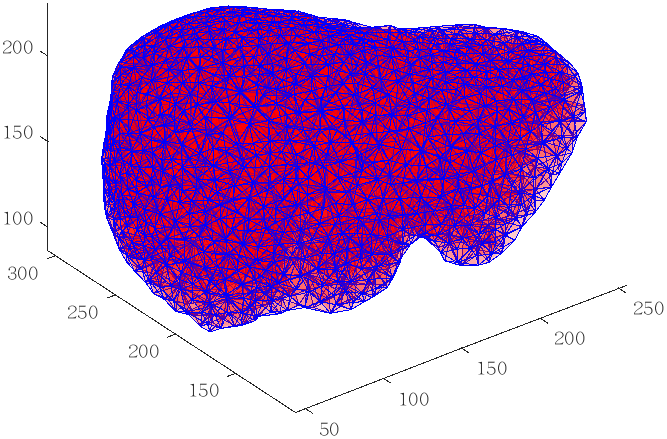}
        \caption{Surface Reconstruction}
    \end{subfigure}
    \caption{Pipeline of data coarsening process: (a) Surface Down-Sample, (b) Surface Down-Sampling, and (c) Surface Reconstruction. We take preoperative mesh of liver model 1 as an example.}
    \label{fig: coarsening} 
\end{figure}

\begin{table}[t]
\caption{TRE separated by the category of intraopeartive measurements used.}
    \centering
    \begin{tabular}{c|r|r|r|r|r|r|r|r}
    \toprule
       Data Splits   &  \multicolumn{4}{c}{Edge TRE} & \multicolumn{4}{|c}{Inner TRE} \\ \cmidrule{2-9}
       & wICP & LIBR & V2S & SR-GCN & wICP & LIBR & V2S & SR-GCN \\ \hline
    Surface data only      &   14.066   &  11.666    &  11.640   &   \textbf{4.405} &  11.869    &  10.412    &  /   &   \textbf{4.412}\\ \hline
    Surface + 1 US Plane   &   14.066   &   9.660   &  11.248   &   \textbf{4.389} &  11.869    &  8.758    &  /   &   \textbf{4.402}\\ \hline
    Surface + 2 US Planes  &  14.066    &   8.174   &  11.15   &   \textbf{4.326} &   11.869   &   7.487   &    / &   \textbf{4.338}\\ \hline
    Surface + 3 US Planes  &  14.066    &  7.225    &   11.05  &   \textbf{4.324} &   11.869   &   6.674   &    / &   \textbf{4.345}\\ \hline
    Surface + 16 US Planes &  14.066    &   4.851   &  10.994   &  \textbf{3.292}  &  11.869    &   4.570   &   /  &   \textbf{3.341}\\ \hline
    All Data               &  14.066    &  7.441    &  11.072   &  \textbf{3.289}  &  11.869    &  6.859    &   /  &   \textbf{3.322}\\ 
    \bottomrule
    \end{tabular}
    \label{tab: baseline2}
\end{table}


\begin{figure}
    \centering
    \begin{subfigure}{0.32\linewidth}
        \includegraphics[width=\linewidth]{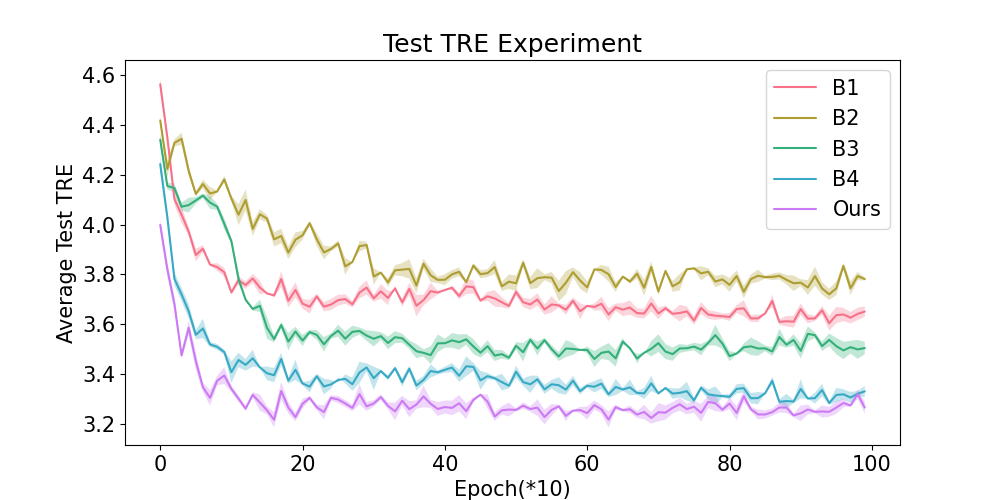}
        \caption{Split 1 Ablative Test TRE Comparison}
    \end{subfigure}
    \hfill
    \begin{subfigure}{0.32\linewidth}
        \includegraphics[width=\linewidth]{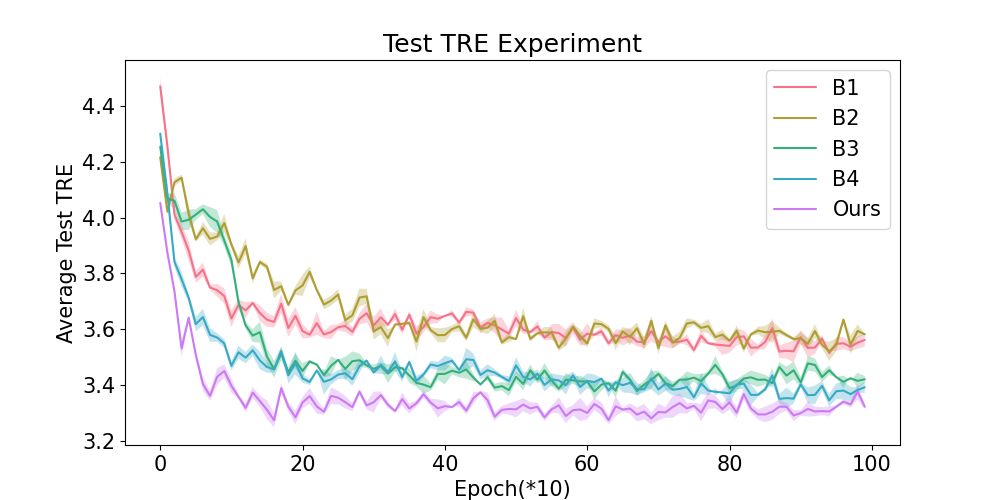}
        \caption{Split 2 Ablative Test TRE Comparison}
    \end{subfigure}
    \hfill
    \begin{subfigure}{0.32\linewidth}
        \includegraphics[width=\linewidth]{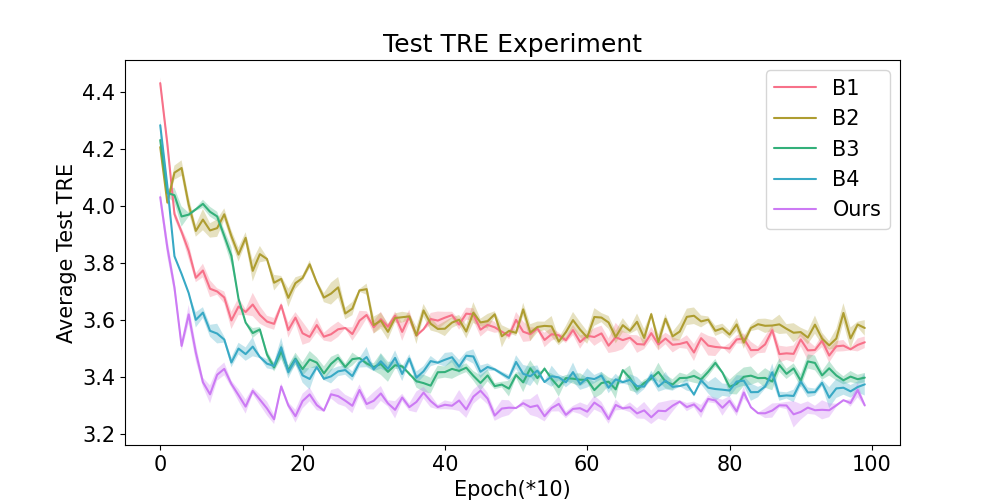}
        \caption{Split 3 Ablative Test TRE Comparison}
    \end{subfigure}\\
    \begin{subfigure}{0.32\linewidth}
        \includegraphics[width=\linewidth]{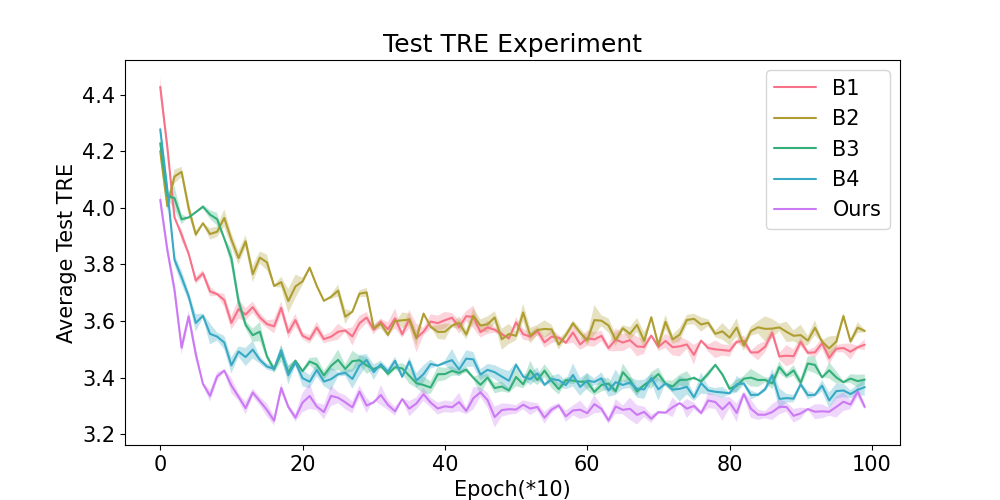}
        \caption{Split 4 Ablative Test TRE Comparison}
    \end{subfigure}
    \hfill
    \begin{subfigure}{0.32\linewidth}
        \includegraphics[width=\linewidth]{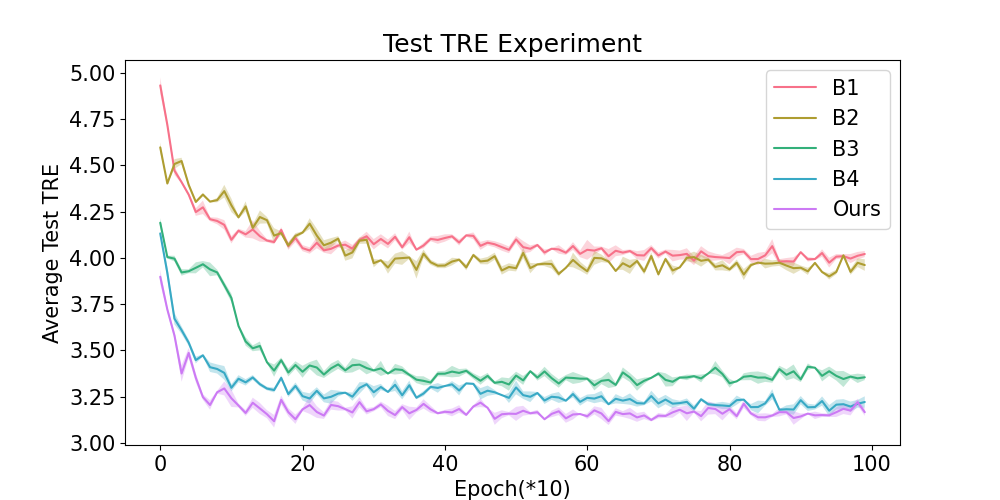}
        \caption{Split 5 Ablative Test TRE Comparison}
    \end{subfigure}
    \hfill
    \begin{subfigure}{0.32\linewidth}
        \includegraphics[width=\linewidth]{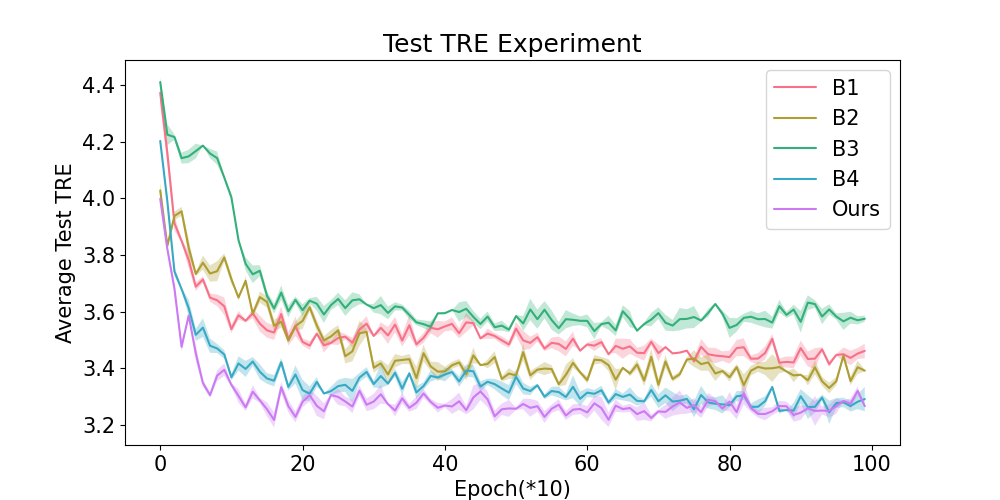}
        \caption{Split 6 Ablative Test TRE Comparison}
    \end{subfigure}
    \caption{Ablative Component Comparison on different data splits: (B1) preoperative mesh as model's  input. (B2) Concatenated preoperative and LIBR-deformed mesh as model's input. (B3) Bipartite Branch Integration on top of (B2). (B4) Sparse loss supervision on top of (B2). (Ours) All components.}
    \label{fig: ablative}
\end{figure}

\end{document}